\documentstyle[prl,aps,twocolumn]{revtex}
\begin{document}
\draft

\twocolumn[\hsize\textwidth\columnwidth\hsize\csname @twocolumnfalse\endcsname

\title{Does the 2D Hubbard Model Really Show $d$-Wave Superconductivity?}
\author{Gang Su$^*$}
\address{Department of Physics, Graduate School, Chinese Academy of Sciences, 
P.O. Box 3908, Beijing 100039, China}
\maketitle

]

In a recent Letter\cite{maier}, Maier et al concluded that they found, based
on the dynamical cluster as well as noncrossing approximations on Green
functions, a stable pure $d$-wave solution with off-diagonal long-range
order (ODLRO) in the intermediate to strong coupling regime for the doped
repulsive 2D Hubbard model, and given an estimation of maximum $T_{c}$ being 
$150K$ for doping $x\approx 0.2$. 

This is a surprising yet too strong statement. The authors in Ref.\cite
{maier} might be unaware of a theorem, proved by Su and Suzuki \cite{su} two
years ago, that the 2D Hubbard model with narrow bands (including next
nearest-neighbor hopping, etc.) does not exhibit $d$-wave pairing long-range
order (LRO) at any nonzero temperature for both repulsive and attractive
Coulomb interactions and for any electron fillings. This rigorous result has
been generalized to cover other models (see, e.g. \cite{noce}). There are
also a few rigorous results showing the absence of $s$-wave or other-wave
pairing LROs for the 2D Hubbard model at finite temperatures\cite{koma,su1}.
It seems that the statement in Ref. \cite{maier} contradicts these
mathematically rigorous results. On the aspect of numerical simulations,
although some quantum Monte Carlo (QMC) calculations show tendencies
favoring $d$-wave superconductivity\cite{scala}, no definite sign of LRO has
been detected, as pointed out by many authors (see, e.g. \cite
{dago,assa,kuro}). Even though the long-tailed enhancements in $d$-wave
pairing correlations near half-filling or $d$-wave-like pairing fluctuations
at low temperatures \cite{scala} were observed in QMC studies, none
concluded from QMC results that the $d$-wave superconductivity with ODLRO is
clearly obtained. In fact, the QMC results, owing to the well-known
limitations, though still controversial, are not incompatible with the
rigorous results. There indeed exist analytic and numerical works (e.g. QMC) 
\cite{veil,huss,zhang}, in completely opposite to the statement claimed in
Ref. \cite{maier}, showing that the 2D Hubbard model (with a next
nearest-neighbor hopping integral) does not exhibit any definite sign of $s$%
-wave and $d$-wave superconductivity. Why are the consequences gained in
Ref. \cite{maier} inconsistent with the rigorous results?

Now, let us look at what was really observed in Ref. \cite{maier}. First,
the method used in Ref. \cite{maier} is only approximate. They used the
coarse grained Green function to mimick the true Green function. According
to the definition, as the coarse grid number $N_{c}$ tends to infinity, the
coarse grained Green function recovers the true Green function, whereas in
the realistic presentations they use just the result of $N_{c}=4$ to draw
final conclusions without giving a convincing comparison for different $%
N_{c} $'s, apart from just stating the systematic errors on the order of $%
1/N_{c}^{3}$. Second, one may notice that the statement of the appearance of
a {\it pure} $d$-wave superconductivity with ODLRO comes from such an
observation that the anomalous coarse grained Green function vanishes at the
points $(0,0)$ and $(\pi ,\pi )$ but is finite at $(0,\pi )$ and $(\pi ,0)$.
Despite how realiable the approximate method is, one cannot, just from this
simple observation, conclude that a {\it pure} $d$-wave superconductivity is
uncovered. For example, a mixing state like $d_{x^{2}-y^{2}}+id_{xy}$ cannot
be ruled out. Third, the transition observed in Ref. \cite{maier} was
speculated as a possibility of a Kosterlitz-Thouless (KT) phase transition.
As is well-known, the correlation function in the KT phase decays in a power
law, while the existence of ODLRO implies, as discussed by Yang\cite{yang},
that the correlation function becomes a nonzero constant as the spatial
separation between the pairing operators tends to infinity. In their seminar
article \cite{kt}, Kosterlitz and Thouless also pointed clearly out that 
{\it this type of phase transition cannot occur in a superconductor}. As a
result, one cannot again just from the approximate results conclude the
occurrence of ODLRO in the 2D Hubbard model at low temperatures.

In short, the results presented in Ref. \cite{maier} are not sufficient to
allow for drawing the conclusion like the authors did, at least in a
conventional sense. As a matter of fact, it appears that the transition they
observed might be a kind of topological phase transition with a power law
decay (i.e. possessing a quasi-LRO), not a genuine superconducting
transition because the latter possesses a true LRO. This reinterpretation
makes the observation in Ref. \cite{maier} consistent with the existing
rigorous results, namely, the 2D Huabbrd model does not show $d$-wave
pairing LRO at finite temperatures.

\end{document}